\documentstyle[11pt]{article}

\begin{document}
\title{Twist Boundary Conditions of Quantum Spin Chains 
near the Gaussian Fixed Points}
\author{A. Kitazawa \\
{\it Department of Physics, Tokyo Institute of Technology,} \\
{\it Oh-okayama, Meguro-ku, Tokyo 152, Japan} }

\date{\today}

\maketitle
\begin{abstract}
Duality transformation, which relates a high-temperature phase to 
a low-temperature one, is used exactly to determine the critical 
point for several models (2D Ising, Potts, Ashkin-Teller, 8-vertex), 
as the self dual condition. 
By changing boundary condition, 
numerically we can determine the self-dual(critical) point of 
the Ashkin-Teller(or Gaussian) model. 
This is the first explicit application of the duality 
to the numerical calculation, with the use of boundary conditions.
\end{abstract}
In this short note, we propose a new method to determine the 2D Gaussian 
critical point of quantum spin chains. 
Although the finite-size scaling method is 
a powerful tool to determine the critical point, difficulty may 
occur for some cases. This difficulty comes from the structure of 
scaling operators. 
By changing the boundary condition, we have the other structure of operators. 
Therefore, selecting boundary conditions, we can use the preferable structure to 
determine the critical point. 
The obtained results are summarized in eq.(\ref{mrslt}) and Fig.2.

As an effective theory of the 1D quantum spin systems, 
the following sine-Gordon model (in Euclidean space-time) 
has been studied
\begin{equation}
  S = \frac{1}{2\pi K}\int d\tau dx \left[
    (\partial_{\tau}\phi)^{2} + (\partial_{x}\phi)^{2} \right]
    +\frac{y}{2\pi\alpha^{2}}\int d\tau dx\cos\sqrt{2}\phi,
\label{sg}
\end{equation}
where $\alpha$ is the lattice constant.
The dual field $\theta(\tau,x)$ (in the Gaussian language) is defined as 
\begin{equation}
  \partial_{\tau}\phi(\tau,x) = -\partial_{x}(iK\theta),\hspace{5mm}
  \partial_{x}\phi(\tau,x) = \partial_{\tau}(iK\theta).
\label{dual}
\end{equation}
We make the identification 
$\phi \equiv \phi+\sqrt{2}\pi,\theta \equiv \theta + \sqrt{2}\pi$.
There exists the $U(1)$ symmetry for the field $\theta$ but 
the second term of eq.(\ref{sg}) violates the $U(1)$ symmetry 
for $\phi$. 
For the free field theory, the scaling dimensions of the vertex operators 
$\exp(\pm in\sqrt{2}\theta)$
and $\exp(\pm im\sqrt{2}\phi)$ are $n^{2}/2K$ and 
$Km^{2}/2$, where the integer variables $n$ and $m$ are electric and magnetic 
charges in the Coulomb gas picture.

After the scaling transformation $a\rightarrow e^{dl}a$, we have the 
following renormalization group equations
\[
  \frac{dK^{-1}}{dl} = \frac{1}{8}y^{2},
  \hspace{5mm}
  \frac{dy}{dl} = (2-\frac{K}{2})y.
\]
These are the famous recursion relations of Kosterlitz. 
Up to the first order of $y$, we find that $y$ is an irrelevant field 
for $K>4$ and relevant for $K<4$. 
There is a separatrix $32K^{-1}-8\ln K^{-1}-y^{2}=0$ which separates 
the infrared unstable region from the infrared stable region, 
and on this separatrix, the Berezinskii-Kosterlitz-Thouless transition occurs. 
The 2D Gaussian fixed line lies on $y=0$. 
For $K<4$ and $y \neq 0$, $y$ flows to infinity. For $y>0$, 
$\langle\phi\rangle$ is renormalized to $\pi/\sqrt{2}$ 
as $y\rightarrow +\infty$ and for $y<0$, 
$\langle\phi\rangle\rightarrow 0$ as $y \rightarrow -\infty$.

The infinite two dimensional plane can be mapped to a periodic strip of 
width $L$ by the conformal mapping $w=(L/2\pi)\log z$ ($z=\tau+ix$). 
In the rest of this note, we consider the boundary effect of this strip system.

First let us consider the following 1-D Hamiltonian 
with the periodic boundary condition\cite{cardy}
\begin{equation}
  H = H_{0}+\lambda\int_{0}^{L} dx{\cal{O}}_{1},
\end{equation}
where $H_{0}$ is the fixed point Hamiltonian 
and ${\cal{O}}_{1}(={\cal{O}}_{1}^{\dagger})$ is a scaling operator 
whose scaling dimension is $x_{1}$.
According to Cardy\cite{cardy}, the following finite size dependence of excitation 
energies up to the first order perturbation is obtained
\begin{equation}
  \Delta E_{n} = \frac{2\pi}{L}\left(
    x_{n}+2\pi\lambda C_{n1n}\left(\frac{2\pi}{L}\right)^{x_{1}-2}
    + \cdots \right),
\label{rel}
\end{equation}
where $L$ is the length of the system, $x_{n}$ is 
the scaling dimension of the operator ${\cal{O}}_{n}$. 
And $C_{n1n}$ is the operator product expansion (OPE) 
coefficient of operators ${\cal{O}}_{n}$ and ${\cal{O}}_{1}$ as
\begin{equation}
{\cal{O}}_{1}(z,\bar{z}){\cal{O}}_{n}(0,0) = 
    C_{n1n} \left( \frac{\alpha}{z}\right)^{h_{1}}
  \left( \frac{\alpha}{\bar{z}}\right)^{\bar{h}_{1}}
  {\cal{O}}_{n}(0,0)
  +\cdots,
\label{fss}
\end{equation}
in which $h_{1}$ and $\bar{h}_{1}$ are the conformal weights of ${\cal{O}}_{1}$ 
($x_{1}=h_{1}+\bar{h}_{1}$).
From eq.(\ref{rel}), we have the following RG equation
\[
  \frac{d\lambda}{d\ln L} = (x_{1}-2)\lambda.
\]
When $x_{1}<2$(relevant), the second order phase transition occurs 
at $\lambda=0$, 
whereas $x_{1}>2$(irrelevant), the second term in eq.(\ref{rel}) is the 
finite size corrections of the excitation energies of the critical systems. 
Up to the first order perturbation theory, we find that 
at the point $\lambda=0$ 
the scaled gap $L\Delta E_{n}$ does not 
depend on the system size and the scaled gaps for several $L$ 
cross linearly at $\lambda=0$.

On the other hand, when the OPE coefficient $C_{n1n}$ becomes zero 
in some reason, the above argument is insufficient and we must consider 
the second order term of $\lambda$ in eq.(\ref{rel}). 
In this case, the scaled gap $L\Delta E_{n}$ may have an extremum 
at the point $\lambda=0$. In practice, this is not 
a preferable thing, because the point of extremum 
is sensitive to finite size corrections of irrelevant operators 
such as $L_{-2}\bar{L}_{-2}${\bf 1} ($x=4$).

In the sine-Gordon model (\ref{sg}), we substitute the operator 
$\sqrt{2}\cos\sqrt{2}\phi$ for ${\cal{O}}_{1}$. 
In this case, there is no operator ${\cal{O}}_{n}$ 
with a nonzero value of 
$\langle{\cal{O}}_{n}^{\dagger}(z_{1}){\cal{O}}_{1}(z_{2}){\cal{O}}_{n}(z_{3})\rangle$
(which is related with the charge neutrality conditions in the Coulomb gas picture.
Note that the operators $e^{\pm i\phi/\sqrt{2}}$ are not allowed.) 
Thus the OPE coefficient in eq.(\ref{rel}) is zero.
This indicates that we cannot expect the simple behavior of 
the finite size scaling method. In addition, for the irrelevant scaling field 
($x>2$), the system is in the massless phase, so 
the finite size scaling method does not work to determine the fixed points. 

If we put artificially half magnetic charges $m=\pm1/2$ in the system, 
the OPE relations are 
\begin{eqnarray}
{\cal{O}}_{1}(z,\bar{z}){\cal{O}}_{1/2}^{e}(0,0) &=& 
    \frac{\sqrt{2}}{2}\left( \frac{\alpha}{z}\right)^{K/4}
  \left( \frac{\alpha}{\bar{z}}\right)^{K/4}
  {\cal{O}}_{1/2}^{e}(0,0)
  +\cdots, \nonumber \\
{\cal{O}}_{1}(z,\bar{z}){\cal{O}}_{1/2}^{o}(0,0) &=& 
  -\frac{\sqrt{2}}{2}\left( \frac{\alpha}{z}\right)^{K/4}
  \left( \frac{\alpha}{\bar{z}}\right)^{K/4}
  {\cal{O}}_{1/2}^{o}(0,0)
  +\cdots, \nonumber \\
\label{half}
\end{eqnarray}
where 
\begin{eqnarray}
  {\cal{O}}_{1}&=&\sqrt{2}\cos\sqrt{2}\phi,
  \nonumber \\
  {\cal{O}}_{1/2}^{e}&=&\sqrt{2}\cos\frac{1}{\sqrt{2}}\phi,
  \\
  {\cal{O}}_{1/2}^{o}&=&\sqrt{2}\sin\frac{1}{\sqrt{2}}\phi,
  \nonumber 
\end{eqnarray}
and there are non-zero OPE coefficients in eq.(\ref{rel}). 

For a physical example with half-odd magnetic charges, 
Alcaraz, Barber, and Batchelor\cite{alcaraz} considered 
the $S=1/2$ XXZ spin chain using Bethe ansatz 
\[
  H = -\sum_{j}^{L}[S_{j}^{x}S_{j+1}^{x}+S_{j}^{y}S_{j+1}^{y}
    +\Delta S_{j}^{z}S_{j+1}^{z}],
\]
with twisted boundary conditions
$S_{L+1}^{x}\pm iS_{L+1}^{y} = e^{\pm i\Phi}(S_{1}^{x}\pm iS_{1}^{y})$, 
$S_{L+1}^{z}=S_{1}^{z}$. 
When $\Phi=0$, this model corresponds to the Gaussian model 
with $K=\pi/\arccos(\Delta)$, $-1<\Delta<1$. 
According to their numerical results, 
the twisted boundary conditions change the electric and magnetic charges as 
\begin{eqnarray}
  n &\rightarrow& n, \nonumber \\
  m &\rightarrow& m+\frac{\Phi}{2\pi}. 
\label{charge}
\end{eqnarray}
Hence when the twist angle $\Phi$ is $\pi$, half odd integer magnetic 
charges appear. 
Recently Fukui and Kawakami\cite{FK} studied this model analytically 
and their results are consistent with eq.(\ref{charge}).
However, since their studies were based on the integrability, 
the off-critical behaviors were not treated.

To see what happens when the boundary condition is changed 
in the Coulomb gas picture, 
we review the case of the following action\cite{crdyetal}
\begin{equation}
  S = \frac{1}{2\pi}K\int_{-\infty}^{\infty}d\tau\int_{0}^{L}dx
    (\partial_{\mu}\theta)^{2}
    +\frac{\Phi}{\sqrt{2}\pi}K\int_{-\infty}^{\infty} 
    d\tau\partial_{x}\theta(\tau,0).
\label{twist}
\end{equation}
Here we write the action with the field $\theta$ which is dual to $\phi$ and 
we assume the periodic boundary condition $\theta(\tau,L)=\theta(\tau,0)$.
If we transform the field $\theta$ as 
$\theta(\tau,x) \rightarrow \theta(\tau,x) - \Phi x/\sqrt{2}L$, 
then we can eliminate the second term of eq.(\ref{twist}) 
with the additional constant term $\Phi^{2}K/2\pi L$, 
but the boundary condition is changed as 
$\theta(\tau,L)=\theta(\tau,0) - \Phi/\sqrt{2}$, 
which corresponds to the defect line along the imaginary time. 
When $\Phi = 2N\pi$ ($N$ is an integer), this is the periodic boundary 
condition. 

After the dual transformation(\ref{dual}), 
the action(\ref{twist}) is transformed as
\begin{equation}
  S = \frac{1}{2\pi K}\int_{-\infty}^{\infty}d\tau\int_{0}^{L}dx
    (\partial_{\mu}\phi)^{2}
    +i\sqrt{2}\left(\frac{\Phi}{2\pi}\right) \int_{-\infty}^{\infty} 
    d\tau\partial_{\tau}\phi(\tau,0).
\end{equation}
This shows that there exist magnetic charges $\mp\Phi/2\pi$ 
at $\tau=\pm\infty$. 
Thus we obtain the ground state energy as\cite{alcaraz}
\begin{equation}
  \frac{2\pi}{L}\left( E_{0}(\Phi) -E_{0}(0)\right)
    = \frac{K}{2}\left(\frac{\Phi}{2\pi}\right)^{2} \equiv x_{0}(\Phi),
\label{exc}
\end{equation}
and the conformal anomaly number changes as
\begin{equation}
  c(\Phi) = 1-12x_{0}(\Phi) = 1-6\left(\frac{\Phi}{2\pi}\right)^{2}K.
\label{con}
\end{equation}
We denote the state corresponding to the vertex operator 
$V_{n,m} = e^{i\sqrt{2}n\theta+i\sqrt{2}m\phi}$ as $|n,m\rangle$. 
Since there exists a magnetic charge $\Phi/2\pi$ at $\tau=-\infty$, 
we find the change of this state as
\begin{equation}
  |n,m\rangle_{\Phi} = |n,m+\Phi/2\pi\rangle_{\Phi=0},
\end{equation}
and because there exists a magnetic charge $-\Phi/2\pi$ at $\tau=\infty$, 
the conjugate state is 
\begin{equation}
  _{\Phi}\langle n,m| = {_{0}}\langle n,m+\Phi/2\pi|.
\end{equation}
Hence we obtain\cite{alcaraz}
\begin{equation}
  E_{n,m}(\Phi)-E_{0}(0) = \frac{2\pi}{L}\left( \frac{n^{2}}{2K}
    +\frac{K}{2}\left(m+\frac{\Phi}{2\pi}\right)^{2}\right),
\label{cphi}
\end{equation}
or
\begin{equation}
  E_{n,m}(\Phi)-E_{0}(\Phi) = \frac{2\pi}{L}\left(\frac{n^{2}}{2K}
    +\frac{K}{2}m^{2}
    +Km\frac{\Phi}{2\pi}\right).
\label{df}
\end{equation}
From this equation, we find that the state $|n,0\rangle_{\Phi}$ corresponds to 
$|n,\Phi/2\pi\rangle_{0}$ which has the excitation energy 
$E_{n,0}(\Phi)-E_{0}(\Phi)= E_{n,0}(0)-E_{0}(0)$, 
and the momentum $n\Phi/L$. 

Note that Dotsenko and Fateev\cite{dotse} considered the similar situation 
\begin{equation}
  S = \frac{1}{2\pi K}\int_{-\infty}^{\infty}d\tau\int_{0}^{L}dx
    (\partial_{\mu}\phi)^{2}
    +i\sqrt{2}\left(\frac{\Phi^{'}}{2\pi}\right) \phi(\tau_{0},0),
  \hspace{5mm} (\tau_{0}\rightarrow\infty).
\end{equation}
in which the additional charge exists only at $\tau=\infty$ but not at 
$\tau=-\infty$. 
The change of conformal anomaly number is the same as eq.(\ref{con}), 
if we set $\Phi^{'} = 2\Phi$\cite{crdyetal}.
But the structure of scaling operators is not same 
with the case of eq.(\ref{twist}).
In their case, $|n,m\rangle_{\Phi} =|n,m\rangle_{0}$, and the conjugate state 
changes as
$_{\Phi}\langle n,m| =  {_{0}}\langle n,m+\Phi^{'}/2\pi|$ 
(which is consistent to eq.(\ref{df})), but in (\ref{twist}) 
the conjugate relation does not change. 
This may be only the difference of picture.

In the case of $\Phi = \pi$, we have half odd integer magnetic
charges effectively. 
In this case, 
$|0,-1\rangle_{\pi}$($=|0,-1/2\rangle_{0}$) and 
$|0\rangle_{\pi}$($=|0,1/2\rangle_{0}$) are degenerate 
for free field theory. 
Introducing the perturbation term of eq.(\ref{sg}) and using the first 
order perturbation theory, 
we obtain the hybridized states
\begin{equation}
  |\psi_{1}\rangle_{\pi} = \frac{1}{\sqrt{2}}
  (|0,-1\rangle_{\pi}+|0\rangle_{\pi})
\label{estate}
\end{equation}
whose parity is even, and 
\begin{equation}
  |\psi_{2}\rangle_{\pi} = \frac{1}{\sqrt{2} i}
  (|0,-1\rangle_{\pi}-|0\rangle_{\pi})
\label{ostate}
\end{equation}
whose parity is odd.
(Note that only when $\Phi=0$ and $\pi$, parity is a good quantum number.) 
Using the OPE(\ref{half}), we obtain the finite size 
dependence of energy up to the first order perturbation as
\begin{eqnarray}
  E_{1}(\pi)-E_{0}(0) &=& \frac{2\pi}{L}
    \left( \frac{K}{8}+2\pi\lambda \frac{\sqrt{2}}{2}
    \left(\frac{2\pi}{L}\right)^{K/2-2}+\cdots \right),
  \nonumber \\
  E_{2}(\pi)-E_{0}(0) &=& \frac{2\pi}{L}
    \left( \frac{K}{8}-2\pi\lambda \frac{\sqrt{2}}{2}
    \left(\frac{2\pi}{L}\right)^{K/2-2}+\cdots \right).
\label{mrslt}
\end{eqnarray}
Thus we find that the energy eigenvalues of these states cross linearly 
at $\lambda=0$. 

In this stage we consider the symmetry of the states(\ref{estate}), (\ref{ostate}).
In the Ashkin-Teller language, the half magnetic charge operator 
$\sqrt{2}\cos\phi/\sqrt{2}$ ($\sqrt{2}\sin\phi/\sqrt{2}$) 
corresponds to the operator 
$\tilde{P} = \sigma^{1}\mu^{2}$ ($\tilde{P}^{*} = \mu^{1}\sigma^{2}$)\cite{KB}.
The sine-Gordon model(\ref{sg}) is invariant under the transformation
\begin{equation}
  \phi \rightarrow \phi + \frac{\pi}{\sqrt{2}},\hspace{5mm}
  \theta \rightarrow \theta,\hspace{3mm}\mbox{and}\hspace{2mm}
  y \rightarrow -y,
\label{dlty} 
\end{equation}
and the operators $\sqrt{2}\cos\phi/\sqrt{2}$ and $\sqrt{2}\sin\phi/\sqrt{2}$ 
are transformed as
\begin{eqnarray}
  \sqrt{2}\cos\phi/\sqrt{2} &\rightarrow& -\sqrt{2}\sin\phi/\sqrt{2},
  \nonumber \\
  \sqrt{2}\sin\phi/\sqrt{2} &\rightarrow& \sqrt{2}\cos\phi/\sqrt{2},
\end{eqnarray}
thus at the point $y=0$ the system has the self-duality\cite{KKd}. 

To verify the above things numerically, we study the following $S=1$ 
quantum spin chain,
\begin{equation}
    H = \sum_{j=1}^{L}(1-\delta(-1)^{j})(S_{j}^{x}S_{j+1}^{x}+S_{j}^{y}S_{j+1}^{y}
    +\Delta S_{j}^{z}S_{j+1}^{z}).
\end{equation}
The effective action of 
this model is described as eq.(\ref{sg}). 
The whole phase diagram was shown in ref.\cite{let}. The transition between 
the dimer and the Haldane phases is of the 2D Gaussian type. 
Using the Lanczos method, we calculate energy eigenvalues of finite systems
($L=8,10,12,14$). 
Figure 1 shows the scaled gap behavior of $L=10,12,14$ systems 
with the periodic boundary condition for $\Delta=0.5$. 
We can see a minimum of the scaled gap. 
In Fig.2, we show two low lying energies of the subspace $\sum S^{z}=0$ with 
the boundary condition $S_{L+1}^{x}=-S_{1}^{x}$, $S_{L+1}^{y}=-S_{1}^{y}$,
$S_{L+1}^{z}=S_{1}^{z}$, which correspond to $E_{1}(\pi)$ and $E_{2}(\pi)$. 
We see the expected behavior(\ref{mrslt}) for this twisted boundary condition. 
The obtained Gaussian fixed points agree with those obtained by the other 
method\cite{NK}. 
The conformal anomaly number is calculated as $c=0.998$ for the periodic boundary 
condition and $c(\pi)=-3.185$ for the $\Phi=\pi$ twisted boundary condition. 
In table 1, we show some extrapolated scaling dimensions. 
These numerical values are consistent with 
eqs.(\ref{exc}), (\ref{con}), (\ref{cphi}), (\ref{mrslt}).
With this method, we can also determine the Gaussian fixed line 
in the massless XY phase\cite{KN} 
and apply to the $S=1$ spin chains 
with the single ion anisotropy\cite{solyom,SZ}.

Lastly we remark the case of the following sine-Gordon model, 
\[
  S = \frac{1}{2\pi K}\int d\tau dx \left[
    (\partial_{\tau}\phi)^{2} + (\partial_{x}\phi)^{2} \right]
    +\frac{y}{2\pi\alpha^{2}}\int d\tau dx\cos\sqrt{8}\phi
\]
with the same operator structure of (\ref{sg}). 
In this case, the three point function 
$\langle e^{\pm i\sqrt{2}\phi(z_{1})}\sqrt{2}\cos\sqrt{8}\phi(z_{2})e^{\pm i\sqrt{2}\phi(z_{3})}\rangle_{0}$
is not zero, so it is enough to consider the periodic boundary condition. 
The transformation corresponding to (\ref{dlty}) is
\[
   \phi \rightarrow \phi + \frac{\pi}{\sqrt{8}},\hspace{5mm}
  \theta \rightarrow \theta,\hspace{3mm}\mbox{and}\hspace{2mm}
  y \rightarrow -y. 
\]
Nomura and Okamoto\cite{NnO} applied the crossing of excitations to determine 
the Gaussian fixed line, 
based on the RG analysis of Giamarchi and Schulz\cite{GS}.

The author thanks K. Nomura for illuminating discussions and critical reading. 
He also acknowledges K. Okamoto for useful discussions. 
The computation in this work has been done 
using the facilities of the Supercomputer Center, 
Institute for Solid State Physics, University of Tokyo.

\begin{table}[b]
\begin{tabular}{c|c|c|c}\hline
  scaling dimension & $x_{1,0}=1/2K$ & $x_{0,1}=K/2$ & $x_{0,1/2}(=x_{0}(\pi))$ \\ \hline
    & $0.1786$ & $1.410$ & $0.3497$  \\ \hline
  $K$ & $2.799$ & $2.819$ & $2.798$   \\
\hline
\end{tabular}
\caption{Scaling dimensions at the critical point 
$\Delta = 0.5$, $\delta=0.2524$. Here we extrapolated the corrections from 
the irrelevant field $L_{2}\bar{L}_{2}${\bf 1} ($x=4$).}
\end{table}

\clearpage

\noindent {\large {\bf Figure captions}}

\vspace{1cm}

\noindent{\bf Fig.1}: The scaled gap behavior of $L=10$($\triangle$), 
$L=12$($\Box$) and $L=14$($\Diamond$) systems 
with the periodic boundary condition for $\Delta=0.5$.

\vspace{5mm}

\noindent{\bf Fig.2}: The low lying energies of the $L=14$ system 
with the $\Phi=\pi$ boundary condition for $\Delta=0.5$. 
Parity even state($E_{1}(\pi)$) is denoted as solid line and 
parity odd state($E_{2}(\pi)$) is denoted as dashed line. 
The crossing point is the critical point and its estimated value is 
$\delta_{c}=0.2524$.

\end{document}